\documentclass[twocolumn,prl,aps,showpacs,floatfix,superscriptaddress]{revtex4}
\usepackage{amsmath,amssymb,graphicx}
\usepackage{color}

\begin{document}

\newcommand{\todo}[1]{\textbf{\textsc{\textcolor{red}{(TODO: #1)}}}}
\newcommand{\OLD}[1]{{\tiny {\bf old:} #1 }}
\newcommand{\NEW}[1]{{ \it #1 }}
\renewcommand{\vec}[1]{\boldsymbol{#1}}
\newcommand{\w}{\omega}

\newcommand{\fcs}{Fe$_{1-x}$Co$_{x}$Si}
\newcommand{\mfs}{Mn$_{1-x}$Fe$_{x}$Si}
\newcommand{\mcs}{Mn$_{1-x}$Co$_{x}$Si}
\newcommand{\cso}{Cu$_{2}$OSeO$_{3}$}

\newcommand{\rxx}{$\rho_{\rm xx}$}
\newcommand{\rxy}{$\rho_{\rm xy}$}
\newcommand{\rxyt}{$\rho_{\rm xy}^{\rm top}$}
\newcommand{\Drxyt}{$\Delta\rho_{\rm xy}^{\rm top}$}
\newcommand{\Sxy}{$\sigma_{\rm xy}$}
\newcommand{\Sxya}{$\sigma_{\rm xy}^A$}

\newcommand{\bco}{$B_{\rm c1}$}
\newcommand{\bct}{$B_{\rm c2}$}
\newcommand{\bao}{$B_{\rm A1}$}
\newcommand{\bat}{$B_{\rm A2}$}
\newcommand{\beff}{$B^{\rm eff}$}

\newcommand{\btr}{$B^{\rm tr}$}

\newcommand{\tc}{$T_{\rm c}$}
\newcommand{\ttr}{$T_{\rm tr}$}

\newcommand{\mb}{$\mu_0\,M/B$}
\newcommand{\dmdb}{$\mu_0\,\mathrm{d}m/\mathrm{d}B$}
\newcommand{\ddmddb}{$\mathrm{\mu_0\Delta}M/\mathrm{\Delta}B$}
\newcommand{\cm}{$\chi_{\rm M}$}
\newcommand{\cac}{$\chi_{\rm ac}$}
\newcommand{\rechi}{${\rm Re}\,\chi_{\rm ac}$}
\newcommand{\imchi}{${\rm Im}\,\chi_{\rm ac}$}

\newcommand{\ozz}{$\langle100\rangle$}
\newcommand{\ooz}{$\langle110\rangle$}
\newcommand{\ooo}{$\langle111\rangle$}
\newcommand{\too}{$\langle211\rangle$}




\title{Real-Space and Reciprocal-Space Berry Phases in the Hall Effect of {\mfs}}

\author{C. Franz}
\affiliation{Physik-Department, Technische Universit\"at M\"unchen,
James-Franck-Stra{\ss}e, D-85748 Garching, Germany}

\author{F. Freimuth}
\affiliation{Institute for Advanced Simulation and Peter Gr\"unberg Institut,
Forschungszentrum J\"ulich and JARA, D-52425 J\"ulich, Germany}

\author{A. Bauer}
\affiliation{Physik-Department, Technische Universit\"at M\"unchen,
James-Franck-Stra{\ss}e, D-85748 Garching, Germany}

\author{R. Ritz}
\affiliation{Physik-Department, Technische Universit\"at M\"unchen,
James-Franck-Stra{\ss}e, D-85748 Garching, Germany}

\author{C. Schnarr}
\affiliation{Physik-Department, Technische Universit\"at M\"unchen,
James-Franck-Stra{\ss}e, D-85748 Garching, Germany}

\author{C. Duvinage}
\affiliation{Physik-Department, Technische Universit\"at M\"unchen,
James-Franck-Stra{\ss}e, D-85748 Garching, Germany}

\author{T. Adams}
\affiliation{Physik-Department, Technische Universit\"at M\"unchen,
James-Franck-Stra{\ss}e, D-85748 Garching, Germany}

\author{S. Bl\"ugel}
\affiliation{Institute for Advanced Simulation and Peter Gr\"unberg Institut,
Forschungszentrum J\"ulich and JARA, D-52425 J\"ulich, Germany}

\author{A. Rosch}
\affiliation{Institute for Theoretical Physics, Universit\"at zu K\"oln, 
Z\"ulpicher Str. 77, D-50937 K\"oln, Germany}

\author{Y. Mokrousov}
\affiliation{Institute for Advanced Simulation and Peter Gr\"unberg Institut,
Forschungszentrum J\"ulich and JARA, D-52425 J\"ulich, Germany}

\author{C. Pfleiderer}
\affiliation{Physik-Department, Technische Universit\"at M\"unchen,
James-Franck-Stra{\ss}e, D-85748 Garching, Germany}

\date{\today}

\begin{abstract}
We report an experimental and computational study of the Hall effect in {\mfs}, as complemented by measurements in {\mcs}, when helimagnetic order is suppressed under substitutional doping. For small $x$ the anomalous Hall effect (AHE) and the topological Hall effect (THE) change sign. Under larger doping the AHE remains small and consistent with the magnetization, while the THE grows by over a factor of ten. Both the sign and the magnitude of the AHE and the THE are in excellent agreement with calculations based on density functional theory. Our study provides the long-sought material-specific microscopic justification, that while the AHE is due to the reciprocal-space Berry curvature, the THE originates in real-space Berry phases. 
\end{abstract}


\vskip2pc

\maketitle


Measurements of the Hall effect in chiral magnets with B20 crystal structure have recently attracted great interest \cite{Neubauer:PRL2009,Ritz:PRB2013,Ritz:Nature2013,Kanazawa:PRL2011,Ishiwata:PRB2011,Porter:PRB2012,Sinha:arXiv:1307.7301}. Due to a hierarchy of energy scales \cite{Landau}, comprising in decreasing strength ferromagnetic exchange, Dzyaloshinsky-Moriya (DM) spin-orbit interactions, and higher order spin-orbit coupling terms, magnetic order in these systems displays generically long-wavelength helical modulations. Under a small applied magnetic field this hierarchy of energy scales stabilizes a skyrmion lattice phase (SLP) in the vicinity of the magnetic transition temperature, i.e., a lattice composed of topologically non-trivial whirls of the magnetization \cite{Muehlbauer:Science2009,Muenzer:PRB2010,Pfleiderer:JPCM2010,Yu:Nature2010,Yu:NatureMaterials2011,Seki:Science2012,Adams:PRL2012,Milde:Science2013}. The Hall effect, which has been studied most extensively in MnSi \cite{Lee:PRB2007,Neubauer:PhysicaB2009,Neubauer:PRL2009,Lee:PRL2009,Ritz:Nature2013,Ritz:PRB2013}, displays thereby three contributions, notably an ordinary Hall effect (OHE), an anomalous Hall effect (AHE) related to the uniform magnetization, and an additional topological Hall effect (THE) in the SLP due to the non-trivial topology of the spin order. 

It was only recently noticed that the THE and AHE represent the real- and reciprocal-space limits of generalised phase-space Berry phases of the conduction electrons, respectively. First principles calculations in MnSi suggest that these phase-space Berry phases account quantitatively for the DM interaction and may even give rise to an electric charge of the skyrmions \cite{Freimuth:arXiv:1307.8085,Freimuth:arXiv:1308.5983}. However, so far perhaps most spectacular because of the experimental evidence is the notion that the non-trivial topological winding of skyrmions gives rise to Berry phases in real space that may be viewed as an emergent magnetic field $B^{\rm eff}=\Phi_0 \Phi$ of one flux quantum ($\Phi_0=h/e$) times the winding number $\Phi=-1$ per skyrmion \cite{Neubauer:PRL2009}. The same mechanism also leads to large spin transfer torques in MnSi \cite{Jonietz:Science2010,Schulz:NaturePhysics2012} and FeGe at ultralow current densities. In turn, a very large THE in MnGe \cite{Kanazawa:PRL2011} and SrFeO$_3$ \cite{Ishiwata:PRB2011} has fuelled speculations that the emergent fields may even approach the quantum limit.  

Despite this wide range of interest, the account of Berry phases in the Hall effect has been essentially phenomenological, in particular for the THE, while a material-specific microscopic justification has been missing. This situation is aggravated by the microscopic sensitivity of the THE to at least three factors: (i) details of the Fermi surface topology, (ii) differences of the average charge carrier life time on each Fermi surface sheet, and (iii) a breakdown of the adiabatic approximation due to spin-flip scattering and possible mixtures of real- and reciprocal-space Berry phases when spin-orbit coupling becomes comparable to the exchange splitting \cite{Ritz:PRB2013}. 

\begin{figure*}
\includegraphics[width=1.0\linewidth,clip=]{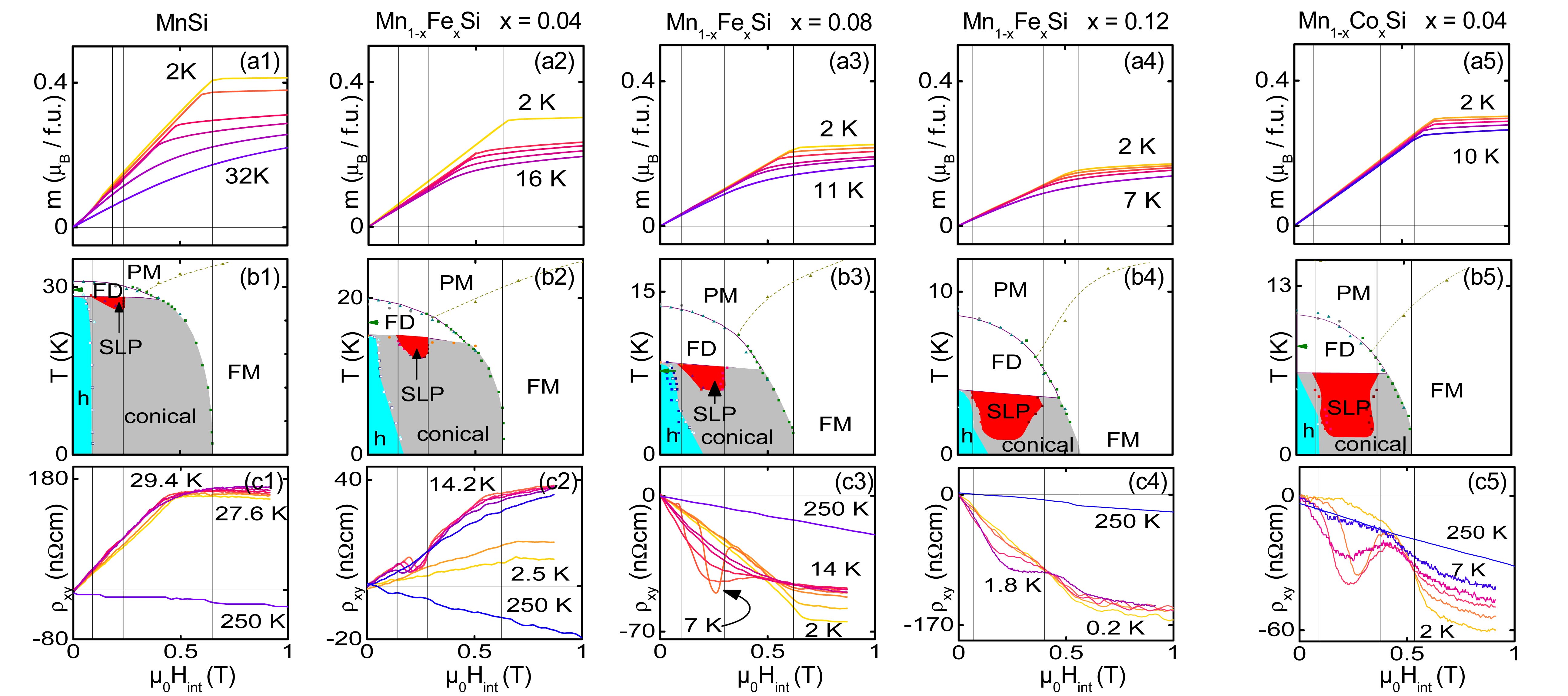}
\caption{(Color online) Doping dependence of $m$, the magnetic phase diagram and {\rxy} of {\mfs} and {\mcs} at selected $x$. Panels (a1) through (a5): Magnetization at selected temperatures in the vicinity of the helimagnetic transition. Panels (b1) through (b5): Magnetic phase diagrams as inferred from the susceptibility calculated numerically from the magnetization (see also Ref.\,\cite{Bauer:PRB2010}). Panels (c1) through (c5): Typical Hall resistivity at selected temperatures (see \cite{SOM} for detailed list of temperatures). At low temperatures the Hall resistivity is dominated by the AHE, which tracks qualitatively the magnetisation. Between $x=0.04$ and 0.08 the sign of this AHE changes from positive to negative. i.e., it assumes qualitatively the shape of a mirror image of $m$. In the field range of the skyrmion lattice phase (SLP), marked by vertical lines, a THE exists in the form of an additional contribution directed downwards for all $x\geq0.04$. For $x=0$ the THE points upwards (cf Fig.\,2 in Ref.\,\cite{Neubauer:PRL2009}), which cannot be resolved on the scale used here, i.e., the THE also changes sign as a function of $x$.
}
\label{figure1}
\end{figure*} 

In this Letter we report a combined experimental and theoretical study of the Hall effect in {\mfs}, supported by complementary Hall data in {\mcs}. As our main results we find for small $x$ a change of sign of both the AHE and THE, however, at slightly different compositions. For larger $x$ we find that the AHE is small consistent with the magnetization, while the magnitude of the THE grows and exceeds that of pure MnSi by over a factor of ten. Using density functional theory we are able to account for both, the magnitude as well as the sign of the THE and AHE observed experimentally. The doping dependence can thereby be related to changes in the ordinary Hall conductivity and a redistribution of $d$-states at the Fermi energy as discussed below. Taken together our study provides the long-sought microscopic justification for the  phenomenological description of the THE and AHE as the real- and reciprocal-space limits of general phase-space Berry phases, with the additional surprise that this occurs in the same complex material.

\begin{figure}
\includegraphics[width=1.0\linewidth,clip=]{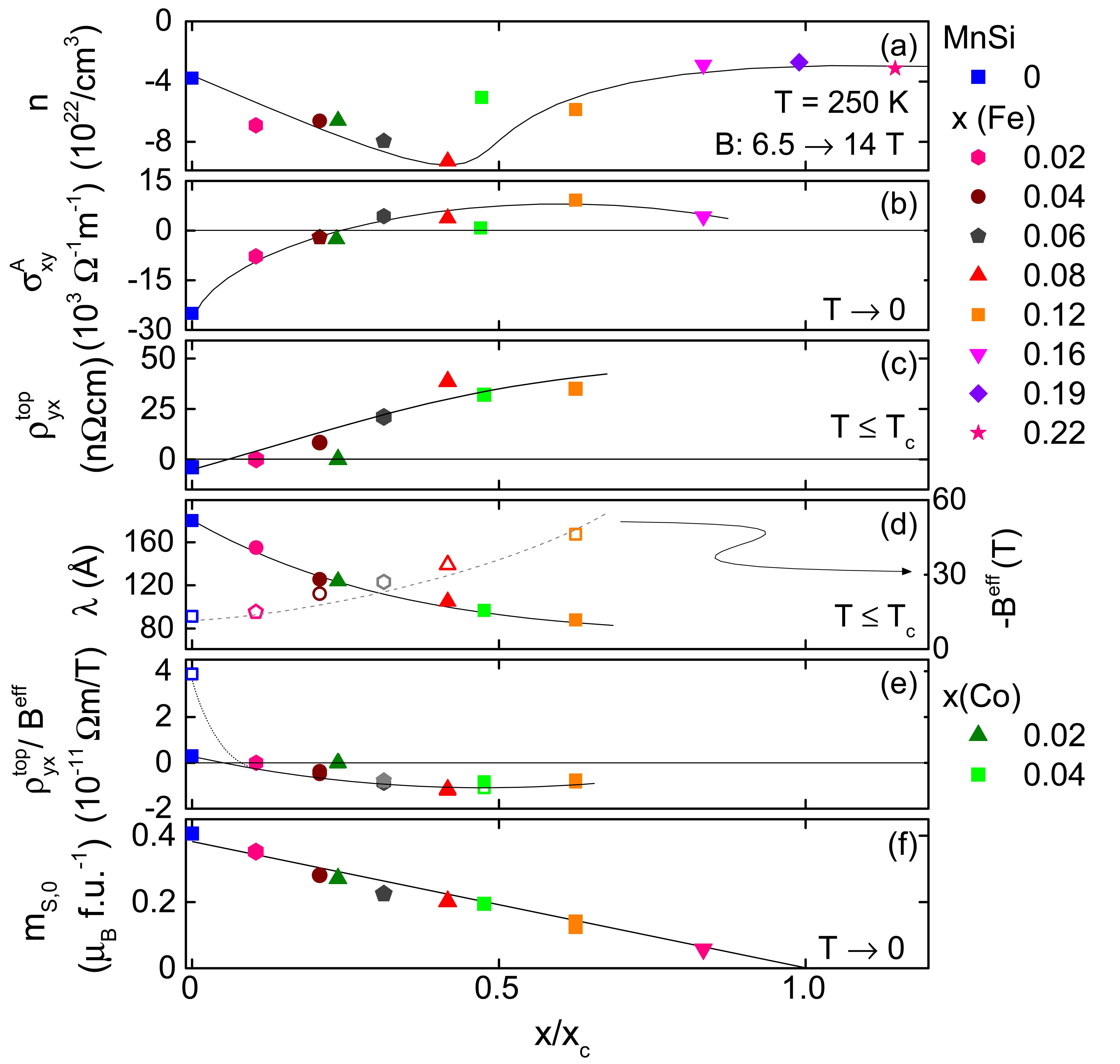}
\caption{(Color online) Main features characterizing the variation of the Hall effects in {\mfs} and {\mcs} as a function of normalized composition $x/x_c$. See text for details.}
\label{figure2}
\end{figure}

For a summary of the experimental methods, which follow the procedures reported in Refs.\,\cite{Ritz:PRB2013,Ritz:Nature2013,Bauer:PRB2010,Bauer:PRB12,Bauer:PRL2013}, we refer to the supplementary information \cite{SOM}. Shown in Fig.\,\ref{figure1}, is an overview of typical magnetisation, and Hall resistivity data as well as magnetic phase diagrams (a detailed list of the temperature values and  parameters studied is part of the supplementary information \cite{SOM}). The evolution of the magnetization $m$ at selected temperatures and selected Fe and Co concentrations is shown in panels (a1) through (a5). At the lowest temperatures $m$ changes from an almost linear increase to being almost field independent when going from below to above {\bct}. With increasing $x$ the magnetization at the lowest temperatures decreases, characteristic of a decrease of the ordered moment (see also Fig.\,\ref{figure2}\,(f) below). The susceptibility calculated from $m$ compares well with the ac susceptibility (both not shown), thus permitting to deduce the magnetic phase diagrams as depicted in Fig.\,\ref{figure1}, panels (b1) through (b5) (cf. Ref.\,\cite{Bauer:PRB2010}). Here the usual magnetic phases are distinguished following accurately previous work \cite{Bauer:PRB2010,Bauer:PRB12,Bauer:PRL2013}, notably paramagnetism (PM), helimagnetic order (h), skyrmion lattice phase (SLP), conical order, field-induced ferromagnetism (FM) and the fluctuation-disorder cross-over regime (FD) \cite{Janoschek:PRB2013}. 

The Hall resistivity {\rxy}, shown in Fig.\,\ref{figure1}, panels (c1) through (c5), displays considerable variations. To permit direct comparison of $m$ with {\rxy} all data have been corrected for the effects of demagnetizing fields. We begin at high temperatures, where {\rxy} is essentially linear and dominated by the OHE without much change as a function of $x$. With decreasing temperature an additional contribution emerges in striking similarity with $m$, the AHE. As a key observation the sign of the AHE changes from positive to negative between $x=0.04$ and 0.06 under Fe-doping (in Fig.\,\ref{figure1} it qualitatively resembles the mirror image of $m$ for $x\geq0.08$; this AHE changes sign between $x=0.02$ and 0.04 under Co-doping). Finally, a third contribution in {\rxy} on top of the OHE and AHE may be distinguished, which for pure MnSi and low concentrations is strictly confined to the SLP. The relevant phase boundaries of the SLP inferred from the susceptibility are marked by vertical lines in Fig.\,\ref{figure1}. For all Fe concentrations $x\geq 0.04$ (under Co doping $x\geq0.02$ ) this THE is negative (downwards) (cf Fig.\,\ref{figure1}\,(c2) to (c5)). In contrast, for the pure compound the THE is positive (upwards) as shown in great detail in Refs\,\cite{Neubauer:PRL2009,Ritz:PRB2013,Ritz:Nature2013}, where the THE for $x=0$ cannot be resolved on the scale of Fig.\,\ref{figure1}\,(c1). Hence, the THE changes sign as a function of $x$, however, at a smaller composition than the AHE. 

As the scope of our paper is the microscopic underpinning of the AHE and THE we do not address the properties for $x\to x_c$, where $T_c$ is suppressed to zero. Nonetheless it is interesting to note that the THE is already present in the fluctuation disordered (FD) regime under Fe-doping, e.g. up to $\sim10\,{\rm K}$ for $x=0.08$ where $T_c=8.8\,{\rm K}$ (cf table of temperature values in supplement \cite{SOM}). This behavior is reminiscent of recent high pressure studies in MnSi, where a THE emerges at high pressure above $T_{\rm c}$ \cite{Ritz:Nature2013}. Further, in the FD regime the THE extends over a much larger field range as compared to the SLP; for $x=0.12$ it clearly emerges around $B=0$ at temperatures down to $T\approx0.6\,{\rm K}$, below which it vanishes. 

Summarized in Fig.\,\ref{figure2} are the salient features of all of our experimental data on the OHE, AHE, and THE. We begin by considering charge carrier concentration $n$ inferred from {\rxy} at 250\,K in the range 6 and 14\,T, i.e. far from the temperature and field range of interest and where it is dominated by the OHE. As shown in Fig.\,\ref{figure2}\,(a) $n$ displays only a gradual reduction by a factor of two around $x/x_c\approx0.5$. This suggests strongly, that the electronic structure does not change radically under doping. Unfortunately it is not possible to obtain any additional information on the OHE at low temperatures and fields, since the AHE and THE are large. 

To capture the AHE we consider the Hall conductivity $\sigma_{\rm xy}=\rho_{\rm yx}/(\rho_{\rm xx}^2+\rho_{\rm xy}^2)$ and determine the anomalous contribution {\Sxya} by extrapolating {\Sxy} from $B>B_{\rm c2}$ to zero field, i.e., we extrapolate from the field-polarized state to zero field. {\Sxya} obtained this way increases from a large negative value at $x=0$, changes sign between $x=0.04$ and 0.06 and approaches a small positive value for large $x$ as shown in Fig.\,\ref{figure2}\,(b). We note that the AHE is large in {\rxy} even for low temperatures (cf. Fig.\,\ref{figure1}, panels (c2) through(c5)), because the residual resistivity increases under doping and approaches $\sim80\,\mu\Omega{\rm cm}$ near $x_c$ \cite{Meingast}. Third, we determined the size of the THE, $\rho_{\rm yx}^{\rm top}$, in the centre of the SLP as described in Refs.\,\cite{Ritz:PRB2013,Ritz:Nature2013}. With increasing doping $\rho_{\rm yx}^{\rm top}$ changes sign between $x=0$ and 0.02, i.e., at a different composition than the AHE. This is followed by an increase of $\rho_{\rm yx}^{\rm top}$ by a factor of ten when increasing $x$ further (Fig.\,\ref{figure2}\,(c)). 

The large increase of {\rxyt} is partly related to the reduction of the helical wavelength $\lambda$ as determined by small-angle neutron scattering (Fig.\,\ref{figure2}\,(d)), where $\lambda$ is found to decrease by a factor of two under doping \cite{Adams:DArbeit,Adams:tbp2013}. Thus, the emergent magnetic field $B^{\rm eff}\sim1/\lambda^2$ increases by a factor of four from -13\,T to about -60\,T over the range of interest as the area per emergent flux quantum $\Phi_0$ decreases. Shown in Fig.\,\ref{figure2}\,(e) is $\rho_{\rm yx}^{\rm top}/B^{\rm eff}$ including an estimate of the effects of finite temperature since the temperature at which the SLP forms decreases with increasing $x$. The estimated zero-temperature values are shown by open symbols. However, the correction is only large for $x=0$, where it was inferred from the pressure dependence \cite{Ritz:PRB2013,metastable}, while it is not even noticeable in the doped samples. In turn the discussion and conclusions presented below do not depend on finite temperature effects.  As a final piece of information needed below we show in Fig.\,\ref{figure2}\,(f) the extrapolated ordered moment in the zero-temperature limit.

\begin{figure}
\includegraphics[width=1.0\linewidth,clip=]{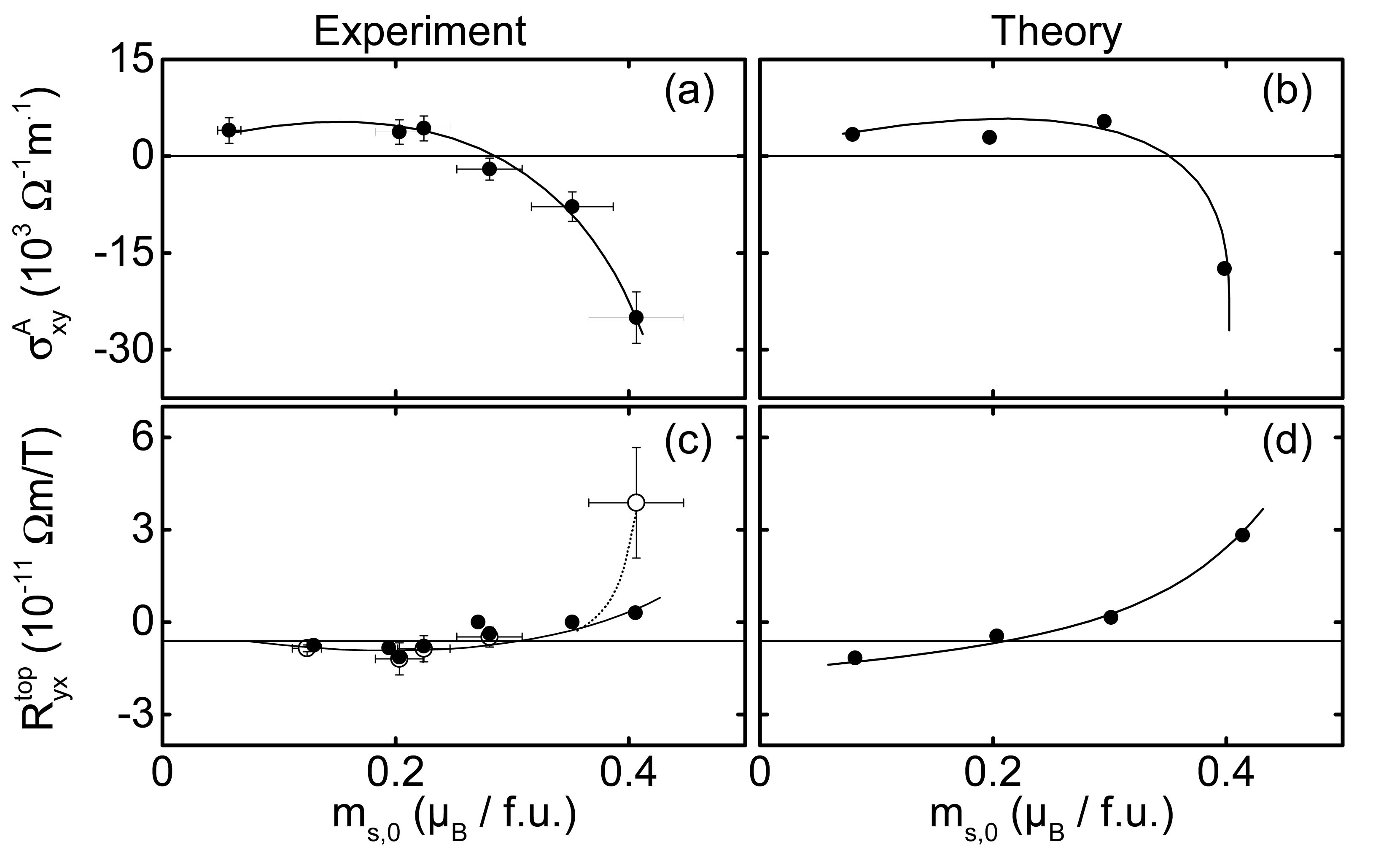}
\caption{Anomalous and topological Hall signal as a function of ordered moment in the zero temperature limit. Open symbols denote the extrapolated value for $T\to0$ as described in the supplement \cite{SOM}. See text for details.
}
\label{figure3}
\end{figure}

In view of the possible microscopic sensitivity of the THE described above \cite{Ritz:PRB2013} we have also considered the Hall effect under Co doping. In a previous study it was found that the temperature versus composition phase diagram, as well as the associated magnetic phase diagrams of {\mfs} and {\mcs}, are remarkably similar \cite{Bauer:PRB2010}. Yet the critical concentration $x_c$ needed to suppress the helimagnetic transition is a factor of two smaller under Co doping. Extending this analogy we find that the sign and magnitude of the AHE and THE as a function of reduced Co concentration, $x/x_c$ also correspond to Fe doping as illustrated in Figs.\,\ref{figure1}\,(a5), \ref{figure1}\,(b5) and \ref{figure1}\,(c5) as well as in Fig.\,\ref{figure2}. This analogy suggests empirically that the detailed relaxation rates are not dominantly responsible for the evolution of the Hall effect under doping. 

For the comparison with experiment we calculated  the electronic structure of collinear ferromagnetic MnSi within the full-potential linearized augmented plane-wave method as implemented in the J\"ulich density functional (DFT) code {\tt FLEUR}. The electronic structure of MnSi was computed within the local density approximation to DFT at the experimental lattice constant, $a=4.558\,{\rm \AA}$. Further, the effect of doping with Fe was taken into account within the virtual crystal approximation (VCA) of MnSi, in terms of a fractional change of the nuclear number at the Mn site proportional to the Fe concentration. Thus, in VCA, adding Fe results in an effective tuning of the electronic structure and Fermi surface topology of pure MnSi, while the details of microscopic electron scattering off Fe impurities are not taken into account. Within VCA, the difference of Co and Fe doping of MnSi arises then from the different nuclear charge of the Mn atoms, and the electronic structure of Mn$_{1-x}$Co$_x$Si  effectively corresponds to the electronic structure of Mn$_{1-2x}$Fe$_{2x}$Si. The approximations made in describing the doping of MnSi with Co and Fe are in turn strongly supported by our experimental results as discussed above.
 
For each concentration $x$ we finally constrained the spin moment to the experimental values shown in Fig.\ 2 (f) (see also \cite{SOM}), where the unconstrained LDA overestimates the spin moments by more than a factor of 2. To compute the Hall conductivities an interpolation based on Wannier functions~\cite{wannier90code} was used. Further computational details are provided in \cite{SOM}.

To confirm that the LDA with constrained spin moment describes the electronic structure correctly, we computed at first the anomalous Hall conductivity (AHC) in ferromagnetic {\mfs}, providing a property that is extremely sensitive to the Fermi surface, and compared it with experiment as shown in Fig.\,\ref{figure3}\,(a) and (b). Namely, we computed the intrinsic contribution to the AHC by evaluating the Berry curvature of the occupied states, neglecting the skew-scattering contribution to the AHE. The latter is suppressed for ferromagnets away from the clean regime, as expected in doped systems like {\mfs}. We have further estimated the extrinsic side-jump contribution to the AHE and find not more than 10\% of the intrinsic values~\cite{Weischenberg}. Hence, the calculated AHC is in excellent agreement with experiment as concerns (i) the magnitude, (ii) the change of sign, and (iii) the reduction of the AHC under Fe doping.

We turn now to the topological Hall constant as determined by Boltzmann transport theory within the constant relaxation time approximation. The latter relies on the assumption that for each spin the relaxation rates of all states at the Fermi surface are the same. Within this approximation the spin-resolved diagonal conductivity is then given by a product of the common relaxation time $\tau_s$ ($s=(\uparrow,\downarrow)$) and a term which is determined by the Fermi surface topology only:  
\begin{equation}
\sigma_{\rm xx}^s = \frac{e^2}{VN}\sum_{\mathbf{k}n}\tau_s\,\delta(E_F-\varepsilon_{\mathbf{k}ns})
\left(  v^{\rm x}_{\mathbf{k}ns} \right)^2
\end{equation}
where $V$ is the unit cell volume, $N$ is the number of $\mathbf{k}$-points in the Brillouin zone, $E_F$ is the Fermi energy, $\varepsilon_{\mathbf{k}ns}$ is the band energy of band $n$ at $\mathbf{k}$, and $v^{\rm x}_{\mathbf{k}ns}$ is the group velocity in ${\rm x}$ direction of this state. We assume that $\tau_s=\alpha \eta_s^{-1}$, where $\eta_s$ is the density of states at $E_F$ for spin $s$ and $\alpha$ is a constant. In turn, the ordinary Hall conductivity at a given magnetic field $B^{\rm z}$ may be expressed as
\begin{eqnarray}
\sigma_{\rm xy}^{{\rm OHE},s}(B^{\rm z})&=&-\frac{e^3B^{\rm z}}{VN}\sum_{\mathbf{k} n}\tau_s^2\,\delta(E_F-\varepsilon_{\mathbf{k}ns})\times \\
&\times& \left[  (v^{\rm x}_{\mathbf{k}ns})^2 m^{\rm yy}_{\mathbf{k}ns} 
- v^{\rm x}_{\mathbf{k}ns}v^{\rm y}_{\mathbf{k}ns} m^{\rm xy}_{\mathbf{k}ns}\right] \nonumber
\end{eqnarray} 
with the inverse effective mass tensor $m^{ij}_{\mathbf{k}ns}=\partial^2\varepsilon_{\mathbf{k}ns}/(\hbar^2\partial k_i\partial k_j)$. The experimentally measured THE has been attributed phenomenologically to the Lorentz force caused by the emergent magnetic field, $B^{\rm eff}=B^{\rm eff,z}$, associated with the spin texture.  This force is opposite for electrons of opposite spin due to the topological charge of the skyrmion, which governs the problem in the space of the magnetization direction. The topological Hall resistivity as approximated by the difference of the OHE for spin-up and spin-down electrons therefore provides a stringent test of this phenomenological Ansatz, namely
\begin{equation}
\rho_{\rm yx}^{\rm top}(B^{\rm eff})=\frac{\sigma_{\rm xy}^{{\rm OHE},\uparrow}(B^{\rm eff}) -  \sigma_{\rm xy}^{{\rm OHE},\downarrow}(B^{\rm eff})}{(\sigma_{\rm xx}^{\uparrow}+\sigma_{\rm xx}^{\downarrow})^2}
\end{equation} 
so that $R^{\rm top}_{\rm yx}=\rho_{\rm yx}^{\rm top}/B^{\rm eff}$ is the topological Hall constant. Within the approximation assumed for the relaxation time,  $R^{\rm top}_{\rm yx}$ does not depend on $\alpha$ and is thus parameter-free. 


Shown in Figs.\,\ref{figure3}\,(c) and (d) are the experimental and calculated dependence of $R^{\rm top}_{\rm yx}$ in Mn$_{1-x}$Fe$_{x}$Si on the ordered magnetic moment. Full symbols correspond to the topological Hall signal at finite temperature, whereas the open symbol corresponds to the estimated zero-temperature value (as explained above this correction is only important in pure MnSi). For pure MnSi the value of about $3.0\times10^{-11}$~$\Omega$m/T compares well to the experimental value of $\sim4.5\times10^{-11}$~$\Omega$m/T. Further, upon doping, both the experimental and theoretical variation of $R^{\rm top}_{\rm xy}$ exhibit a change of sign and a noticeable reduction in magnitude in reasonable agreement. 

According to our calculations all of this can be understood on the basis of the band structure of paramagnetic MnSi in terms of (i) the change of sign of $\sigma_{\rm xy}^{\rm OHE}$ for both spin channels, (ii) a significant decrease in magnitude of $\sigma_{\rm xy}^{{\rm OHE},\uparrow}$ as the Fe concentration increases, and (iii) a redistribution of the $d$-states at the Fermi energy \cite{SOM}. Graphical illustrations highlighting these observations are shown as part of the supplementary information \cite{SOM}. The change of sign in the OHE is only accessible in our calculations but cannot be determined at low temperatures and small fields.

In summary, the excellent agreement between experiment and microscopic calculations of the AHE and THE in {\mfs} reported here provide the long-sought material-specific microscopic justification for the extreme limits of real-space and reciprocal-space Berry phases in a multi-band metal with a complex multi-sheeted Fermi surface. Our study thereby demonstrates the enormous sensitivity of the magnitude of the topological Hall signal to details of the electronic structure, in particular to the Fermi surface topology. Even if the emergent field is very large as in weakly Fe- or Co-doped MnSi, where $B^{\rm eff}\approx-15\,{\rm T}$, the THE may vanish altogether.

We wish to thank P. B\"oni and S. Mayr for helpful discussions and support. CF, AB, RR, CS, CD and TA acknowledge financial support through the TUM Graduate School. Financial support through DFG TRR80 and ERC-AdG (291079 TOPFIT) are gratefully acknowledged. FF and YM were financially supported under project YIG VH-NG-513 of the Helmholtz Gemeinschaft. FF, SB and YM are grateful to the J\"ulich Supercomputing Centre for computational time.


\end{document}


\newcommand{\todo}[1]{\textbf{\textsc{\textcolor{red}{(TODO: #1)}}}}
\newcommand{\OLD}[1]{{\tiny {\bf old:} #1 }}
\newcommand{\NEW}[1]{{ \it #1 }}
\renewcommand{\vec}[1]{\boldsymbol{#1}}
\newcommand{\w}{\omega}

\newcommand{\fcs}{Fe$_{1-x}$Co$_{x}$Si}
\newcommand{\mfs}{Mn$_{1-x}$Fe$_{x}$Si}
\newcommand{\mcs}{Mn$_{1-x}$Co$_{x}$Si}
\newcommand{\cso}{Cu$_{2}$OSeO$_{3}$}

\newcommand{\rxx}{$\rho_{\rm xx}$}
\newcommand{\rxy}{$\rho_{\rm xy}$}
\newcommand{\rxyt}{$\rho_{\rm xy}^{\rm T}$}
\newcommand{\Drxyt}{$\Delta\rho_{\rm xy}^{\rm top}$}
\newcommand{\Sxy}{$\sigma_{\rm xy}$}
\newcommand{\Sxya}{$\sigma_{\rm xy}^A$}

\newcommand{\bco}{$B_{\rm c1}$}
\newcommand{\bct}{$B_{\rm c2}$}
\newcommand{\bao}{$B_{\rm A1}$}
\newcommand{\bat}{$B_{\rm A2}$}
\newcommand{\beff}{$B^{\rm eff}$}

\newcommand{\btr}{$B^{\rm tr}$}

\newcommand{\tc}{$T_{\rm c}$}
\newcommand{\ttr}{$T_{\rm tr}$}

\newcommand{\mb}{$\mu_0\,M/B$}
\newcommand{\dmdb}{$\mu_0\,\mathrm{d}M/\mathrm{d}B$}
\newcommand{\ddmddb}{$\mathrm{\mu_0\Delta}M/\mathrm{\Delta}B$}
\newcommand{\cm}{$\chi_{\rm M}$}
\newcommand{\cac}{$\chi_{\rm ac}$}
\newcommand{\rechi}{${\rm Re}\,\chi_{\rm ac}$}
\newcommand{\imchi}{${\rm Im}\,\chi_{\rm ac}$}

\newcommand{\ozz}{$\langle100\rangle$}
\newcommand{\ooz}{$\langle110\rangle$}
\newcommand{\ooo}{$\langle111\rangle$}
\newcommand{\too}{$\langle211\rangle$}

\makeatletter 
\renewcommand{\thefigure}{S\@arabic\c@figure}
\makeatother
\setcounter{secnumdepth}{2}

\title{Supplementary Material for:\\ Real-Space and Reciprocal-Space Berry-Phases in the Hall-Effect of {\mfs}}

\author{C. Franz}
\affiliation{Physik-Department, Technische Universit\"at M\"unchen,
James-Franck-Stra{\ss}e, D-85748 Garching, Germany}

\author{F. Freimuth}
\affiliation{Institute for Advanced Simulation and Peter Gr\"unberg Institut,
Forschungszentrum J\"ulich and JARA, D-52425 J\"ulich, Germany}

\author{A. Bauer}
\affiliation{Physik-Department, Technische Universit\"at M\"unchen,
James-Franck-Stra{\ss}e, D-85748 Garching, Germany}

\author{R. Ritz}
\affiliation{Physik-Department, Technische Universit\"at M\"unchen,
James-Franck-Stra{\ss}e, D-85748 Garching, Germany}

\author{C. Schnarr}
\affiliation{Physik-Department, Technische Universit\"at M\"unchen,
James-Franck-Stra{\ss}e, D-85748 Garching, Germany}

\author{C. Duvinage}
\affiliation{Physik-Department, Technische Universit\"at M\"unchen,
James-Franck-Stra{\ss}e, D-85748 Garching, Germany}

\author{T. Adams}
\affiliation{Physik-Department, Technische Universit\"at M\"unchen,
James-Franck-Stra{\ss}e, D-85748 Garching, Germany}

\author{S. Bl\"ugel }
\affiliation{Institute for Advanced Simulation and Peter Gr\"unberg Institut,
Forschungszentrum J\"ulich and JARA, D-52425 J\"ulich, Germany}

\author{A. Rosch}
\affiliation{Institute for Theoretical Physics, Universit\"at zu K\"oln, 
Z\"ulpicher Str. 77, D-50937 K\"oln, Germany}

\author{Y. Mokrousov}
\affiliation{Institute for Advanced Simulation and Peter Gr\"unberg Institut,
Forschungszentrum J\"ulich and JARA, D-52425 J\"ulich, Germany}

\author{C. Pfleiderer}
\affiliation{Physik-Department, Technische Universit\"at M\"unchen,
James-Franck-Stra{\ss}e, D-85748 Garching, Germany}

\date{\today}

\begin{abstract}
We report supplementary information on the data shown in the main text, the sample preparation and experimental set-up, and the method used to account for the reduction of the topological Hall signal at finite temperatures. In addition, technical details are described of the calculational procedures.
\end{abstract}


\vskip2pc

\maketitle

\section{Experimental Methods}

\subsection{Data presentation in Fig.\,1}

Summarised in table\,\ref{table1} are all temperature values in Kelvin at which data were recorded for this study, as well as the temperature values of the data shown in Fig.\,1 of the main text. The order in which values are listed in this table corresponds to the order of the curves on the right hand side of each panel from top (yellow) to bottom (blue). In Fig.\,1\,(a1) data reside largely on top of each other and cannot be easily resolved. For detailed information and plots we refer to Refs.\,\cite{Neubauer:PRL2009,Ritz:PRB2013,Ritz:Nature2013}. 

\subsection{Sample Preparation and Set-up}

For our study single crystals of MnSi, {\mfs} ($x=0.02$, 0.04, 0.06, 0.08, 0.12, 0.16, 0.19, 0.22) and {\mcs} ($x=0.02$, 0.04) were grown by optical float-zoning \cite{Neubauer:RSI2011,Bauer:PRB2010}. For information on the magnetisation, susceptibility and specific heat of samples taken from the same ingots we refer to Ref.\,\cite{Bauer:PRB2010,Bauer:thesis,Bauer:PRL2013,Bauer:PRB12}, while small angle neutron scattering will be reported in a forthcoming paper \cite{Adams:tbp2013}. Samples for our transport measurements were oriented by Laue x-ray diffraction, cut with a wire saw, and carefully polished to size. Au wires were spot welded to the samples and the geometry factor determined with an optical light microscope.

\begin{sidewaystable*}
\caption{Summary of the temperatures in Kelvin for which data were recorded and for which data are shown in Fig.\,1 of the main text. Order of values of temperatures for which data are shown in Fig.\,1 of the main text correspond thereby to the curves on the right hand side of each panel from top to bottom. Values of the helimagnetic transition temperature $T_{\rm c}$ were determined as described in Ref.\,\cite{Bauer:PRB2010}.
}
\begin{tabular}{|l|p{2.5cm}|p{2.5cm}|p{2.5cm}|p{2.5cm}|p{2.5cm}|p{2.5cm}|p{0.02cm}|p{2.5cm}|p{2.5cm}|}
\cline{1-7}\cline{9-10} 
& x\,=\,0 & x\,=\,0.02 & x\,=\,0.04 & x\,=\,0.06 & x\,=\,0.08 & x\,=\,0.12 & & x\,=\,0.02 & x\,=\,0.04\\
\cline{1-7}\cline{9-10} 
dopant &  & Fe & Fe &	Fe & Fe & Fe && Co & Co \\
\cline{1-7}\cline{9-10} 
T$_c$ in (K) & 29.5 & 20.2 & 15.1 &	10.3 & 8.8 & 6.0 && 14.9 & 6.7 \\
\cline{1-7}\cline{9-10} 
$m$ all values 	
&2, 4, 8, 12, 16, 20, 24, 26, 27, 28, 29, 30, 31, 32 
&2, 4, 8, 12, 16, 18, 19, 20, 21, 22, 24, 26, 30 
&2, 4, 6, 8, 10, 11, 12, 12.5, 13, 13.5, 14, 14.5 15, 15.5, 16, 16.5 17, 18, 20
&2, 4, 6, 8, 9, 10, 11, 12, 13, 14, 15, 16, 18, 20 
&2, 2.5, 3, 3.5, 4, 4.5, 5, 5.5, 6, 6.5, 7, 7.5, 8, 8.5, 9, 9.5, 10, 10.5, 11
&2. 2.5, 3, 3.5, 4, 4.5, 5, 5.5, 6, 7, 10
&
& 2, 4, 6, 8, 10, 11, 11.5, 12, 12.5, 13, 13.5, 14, 14.5, 15, 16, 20
& 2, 4, 6, 8, 10\\
\cline{1-7}\cline{9-10} 
$m$ as shown
& Fig.\,1\,(a1)
&
& Fig.\,1\,(a2)
&
& Fig.\,1\,(a3)
& Fig.\,1\,(a4) 
&&
& Fig.\,1\,(a5)\\
in Fig.\,1
&2, 16, 26, 30, 32 
&	 
&2, 12.5, 13.5 15, 16 
& 
&2, 4, 6, 8, 9, 11 
&2, 3, 4, 5, 7 
&&
& 2, 4, 6, 8, 10\\
\cline{1-7}\cline{9-10} 
$\rho_{\rm xy}$ all values	
& 4, 10, 25, 50, 75, 100, 150, 200, 250\newline\newline
from 27.6$\to$29.4
in steps of 0.1
&2, 4, 10, 18.4, 18.6, 19, 19.5, 19.7, 19.9, 20.1, 20.3, 20.5, 25, 100, 250
&2.5, 2.6, 4, 5, 10, 12.6, 12.8, 13, 13.1, 13.2, 13.3, 13.4, 13.5, 13.6, 13.8, 14, 14.1, 14.2, 14.4, 14.6, 14.8, 15, 25, 50, 100, 150, 200, 250 
&2.1, 4, 5.2, 8.2, 8.4, 8.6, 8.8, 9, 9.2, 9.4, 9.6, 9.8, 10, 10.2, 10.4, 10.6, 10.8, 11, 11.2, 11.4, 11.8, 12, 12,2, 25, 100, 250 
&0.08, 0.5, 2, 3, 4, 5, 6, 6.5, 7, 7.5, 8, 8.5, 9, 10, 11, 12, 13, 14, 15, 16, 17, 18, 19, 20, 25, 50, 100, 150, 200, 250
&0.2, 0.5, 0.6, 0.8, 1.6, 1.8, 2, 2.2, 2.6, 2.8, 3, 3.2, 3.4, 3.6, 3.8, 4, 5, 6, 7, 8, 10, 25, 50, 50, 75, 100, 125, 150, 175, 200, 225, 250
&
& 4, 10, 12, 12.5, 13, 13.5, 14, 14.5, 15, 15.5, 16, 16.5, 17, 25, 50, 100, 150, 200, 250
& 2, 2.5, 3, 3.5, 4, 4.5, 5, 5.5, 6, 6.5, 7, 10, 25, 50, 100, 150, 200, 250
\\
\cline{1-7}\cline{9-10} 
$\rho_{\rm xy}$ as shown
& Fig.\,1\,(c1)
&
& Fig.\,1\,(c2)
&
& Fig.\,1\,(c3)
& Fig.\,1\,(c4) 
&&
& Fig.\,1\,(c5)\\
in Fig.\,1
&27.6, 27.9, 28.2, 28.5, 28.9, 29.4, 250\newline
see Ref.\,\cite{Neubauer:PRL2009,Ritz:PRB2013,Ritz:Nature2013}\newline
for detailed plots
&
&2.5, 5, 13, 13.4, 13.8, 14.2, 15, 250 
&
&2, 4, 6, 7, 8, 9, 12, 14, 250
&0.2, 0.5, 0.6, 0.8, 1.8, 250
&&
& 2, 3, 4, 5, 7, 250\\
\cline{1-7}\cline{9-10} 
\end{tabular}
\label{table1}
\end{sidewaystable*}

\subsection{Signal Detection and Processing}

The Hall and the longitudinal resistivity, $\rho_{\rm xy}$ and $\rho_{\rm xx}$, respectively, were measured simultaneously with a standard six-terminal phase-sensitive detection system. Low excitation frequencies and low excitation currents were used to avoid parasitic signal pick-up. The Hall signal $\rho_{\rm xy}$  and the longitudinal signal, $\rho_{\rm xx}$, correspond thereby to the antisymmetric or symmetric component of the transverse or longitudinal voltage pick-up in magnetic field sweeps.

\subsection{Effects of Non-Zero Temperature}

In our study we obtained an empirical estimate of the effects of finite temperatures on the topological Hall signal in {\mfs} and {\mcs} with the help of a recent high pressure study in pure MnSi \cite{Ritz:PRB2013,Ritz:Nature2013}. The upshot of these estimates is, that the effects of finite temperatures do not affect the results of our study. In the high pressure study careful measurements of the Hall effect revealed a high sensitivity to the temperature and magnetic field history for pressures between 2\,kbar and roughly 12\,kbar. As the main observation it was thereby found that the topological Hall signal could be stabilised down to the lowest temperatures when field-cooling the sample. In turn this suggested that the skyrmion lattice survived down to the lowest temperatures in a metastable (super-cooled) state. Similar metastable behavior has recently been observed and investigated in {\fcs} by means of small angle neutron scattering and magnetic force microscopy \cite{Muenzer:PRB2010,Milde:Science2013}. Comparison of the Hall signal under field-cooling and zero-field cooling provides finally an estimate of an upper boundary of the topological Hall signal when switching off the effects of temperature.

\begin{figure}
\includegraphics[width=0.85\linewidth,clip=]{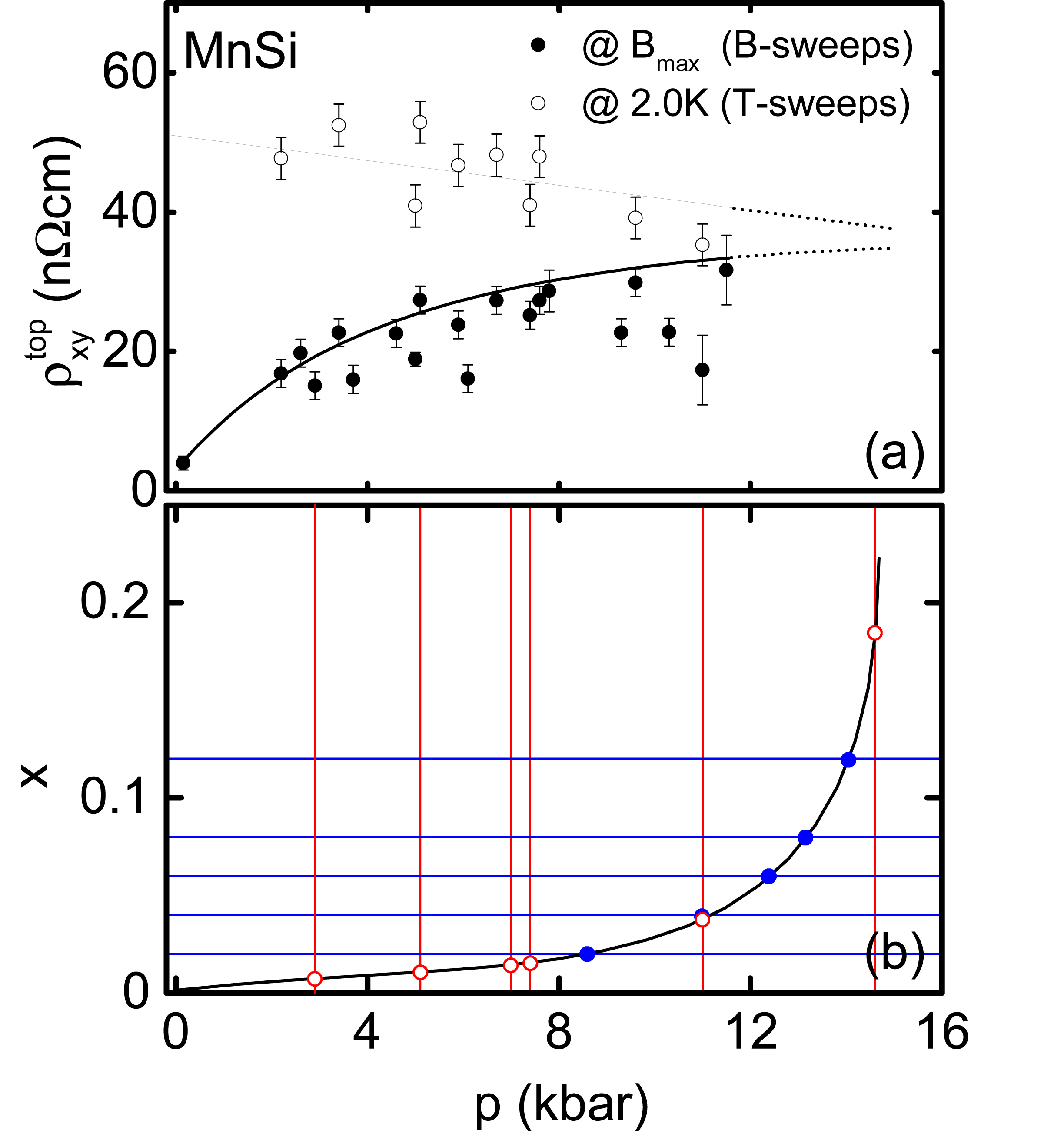}
\caption{(a) Pressure dependence of the topological Hall signal of pure MnSi under field cooling and zero-field cooling (see text for details). Lines are guides to the eye. The plot is a reproduction from data reported, e.g., in Ref.\,\cite{Ritz:PRB2013}. (b) Relationship between doping and pressure with the transition temperature as implicit parameter. Red lines mark pressures at which transition temperatures have been determined in previous work \cite{Ritz:PRB2013}, blue lines mark compositions at which transition temperatures have been determined. 
}
\label{figureS0a}
\end{figure}

\begin{figure}
\includegraphics[width=0.85\linewidth,clip=]{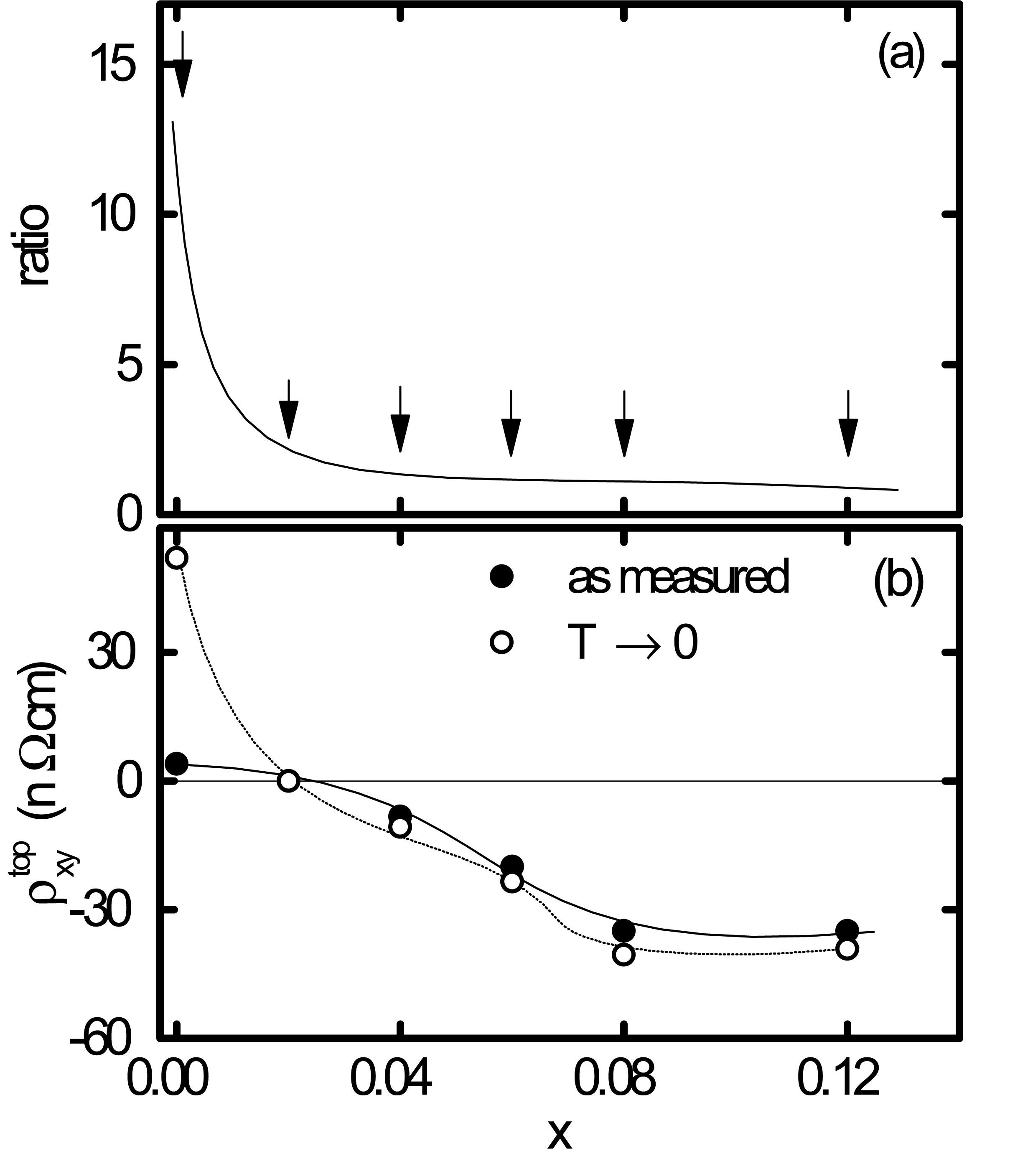}
\caption{(a) Ratio of the topological Hall signal after field-cooling over signal size in the reversible regime at finite temperature. The ratio is shown as a function of composition, where Fig.\,\ref{figureS0a}\,(b) illustrates the pressure - composition relationship. (b) Topological Hall signal, {\rxyt}, as measured (filled symbols) and as corrected by the correction factor shown on panel (a), (open symbols). The correction is only large for the pure compound. 
}
\label{figureS0b}
\end{figure}

Shown in Fig.\,\ref{figureS0a}\,(a) are the high-pressure data recorded in pure MnSi. Open symbols represent the Hall signal under field-cooling down to 2\,K. These data compare with the filled symbols, which represent an estimate of the topological Hall signal recorded in field sweeps of the skyrmion lattice phase. The latter is equivalent to the behaviour under zero-field-cooling and corresponds to the reversible parameter regime of the skyrmion lattice phase. This regime is located in the vicinity of the helimagnetic transition temperature. The lines through the data points in Fig.\,\ref{figureS0a} are guides to the eye. They are the basis for our estimate of the topological signal in the vicinity of $T_c$ when switching off the effects of finite temperature. Notably, the ratio of the upper to the lower line provides a factor, that describes the reduction of the topological Hall effect at the temperature at which the topological Hall effect was measured for the particular pressure.

To translate pressure into an equivalent composition at which the correction factor is applied we show in  Fig.\,\ref{figureS0a}\,(b) the pressure versus composition relationship at which the helimagnetic transition temperatures are the same. Since the transition temperature is an implicit parameter we mark in this figure pressures at which transition temperatures have been recorded in previous studies (e.g. \cite{Ritz:PRB2013}) by red lines and doping compositions at which transition temperatures have been determined by blue lines. Using this relationship we show in Fig.\,\ref{figureS0b}\,(a) the correction factor as a function of Fe content. Applying this factor, we finally arrive at a comparison of the topological Hall resistivity {\rxyt} as measured (filled symbols) with the extrapolated generic value for $T\to0$ when applying the correction factor (open symbols). The correction is thereby only large for pure MnSi, in which the transition temperature is around 30\,K. It is small when the transition temperature becomes reasonably small.

\section{Computational Aspects}
We computed the electronic structure of MnSi within the LDA using atomic coordinates and lattice parameter ($a$=4.558\AA) of the 8 atom unit cell as given in~\cite{MnSi_Pickett}. We set the muffin-tin radii to 2.12$a_0$ for both Mn and Si and the plane-wave cutoff to 3.7$a_0^{-1}$ ($a_0=0.529177$\AA\, is Bohr's radius) and used 10648 k-points in the full Brillouin zone for the self-consistent calculation of the electronic
charge density.

For the virtual crystal potential approximation (VCA) calculations of Mn$_{1-x}$Fe$_{x}$Si we used the same lattice constant and computational parameters as for MnSi. The magnetic moments obtained within LDA do not match the experimental values given in Fig.~2~(f). For example, the unconstrained spin moment of MnSi is 0.94 $\mu_{\rm B}$ per formula unit. Therefore, we constrained the moments in our calculation to the experimental values shown in Fig.~2~(f) of the main text for each concentration $x$. Employing the Wannier90 code~\cite{wannier90code}, we constructed 72 relativistic maximally localized Wannier functions using disentanglement from 120 bands on an 8$\times$8$\times$8 k-mesh~\cite{WannierPaper}. In order to compute the Hall coefficients efficiently we made use of the Wannier interpolation~\cite{ahe_wannier_interpolation,wannier_interpolation} with a 512$\times$512$\times$512 k-mesh. 

\begin{figure}
\includegraphics[width=0.85\linewidth,clip=]{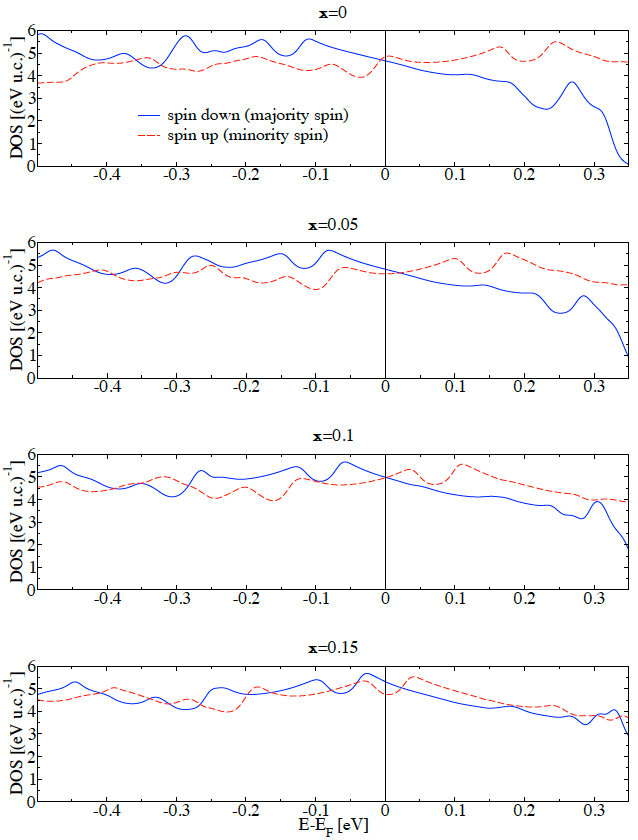}
\caption{Evolution of the spin-resolved DOS with
doping $x$. The doping dependence is dominated by a shift of the Fermi level together with a decrease of the exchange splitting.}
\label{fig_dos_change_by_doping}
\end{figure}

\begin{figure}
\includegraphics[width=0.85\linewidth,clip=]{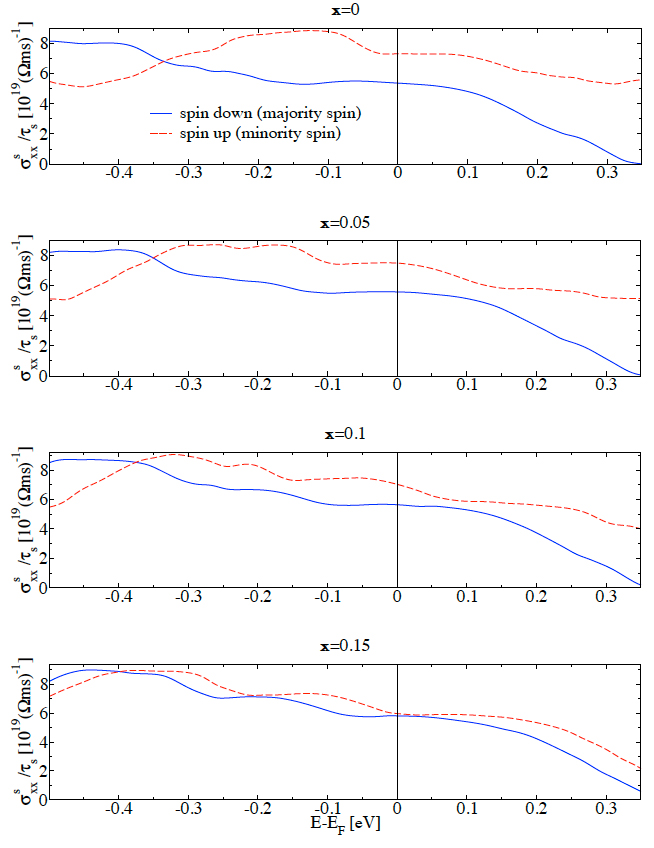}
\caption{Evolution of the spin-resolved normal conductivity with
doping $x$. }
\label{fig_conductivity_change_by_doping}
\end{figure}

\begin{figure}
\includegraphics[width=0.85\linewidth,clip=]{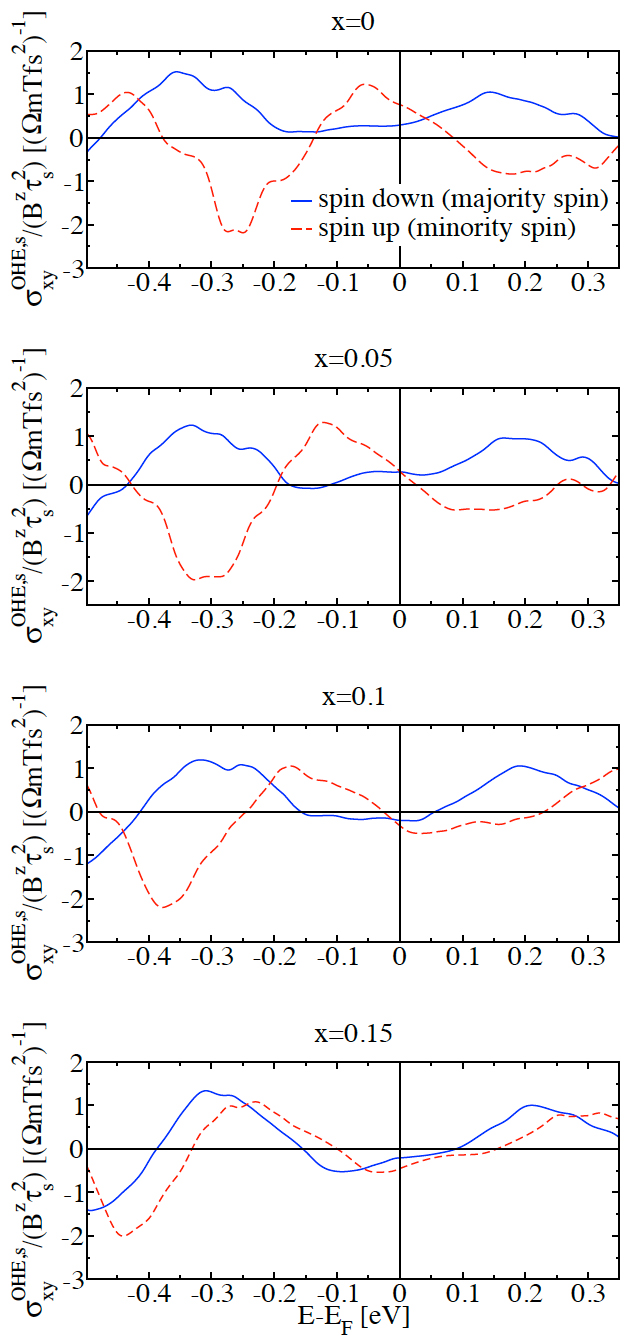}
\caption{Evolution of the spin-resolved ordinary Hall conductivity with
doping $x$. Between $x=0$ and $x=0.1$ both $\sigma_{xy}^{{\rm OHE},\uparrow}$ and $\sigma_{xy}^{{\rm OHE},\downarrow}$ and consequently also $R^{\rm top}_{yx}$ change sign. Additionally, the magnitude of $\sigma_{xy}^{{\rm OHE},\uparrow}$ is reduced considerably.}
\label{fig_ohe_change_by_doping}
\end{figure}

Fig.~3 of the main text shows a strong dependence of the THE on doping. In order to relate this strong dependence to the doping-induced changes of the electronic structure, we show the evolution with doping of the spin-resolved density of states (DOS), the spin-resolved normal conductivity $\sigma^{\rm s}_{\rm xx}$ and the spin-resolved ordinary Hall conductivity $\sigma^{\rm OHE,s}_{\rm xy}$ in Figs.~\ref{fig_dos_change_by_doping}, \ref{fig_conductivity_change_by_doping} and \ref{fig_ohe_change_by_doping}, respectively, where the spin is labelled by s. According to Fig.~\ref{fig_dos_change_by_doping} the evolution of the DOS with doping $x$ is dominated by a shift of the Fermi level together with a decrease of the exchange splitting. The spin-resolved OHE conductivity $\sigma^{\rm OHE,s}_{\rm xy}$ shown in Fig.~\ref{fig_ohe_change_by_doping} exhibits a strong dependence on the doping $x$, while both the DOS and the spin-resolved normal conductivity (shown in Fig.~\ref{fig_conductivity_change_by_doping}) vary only weakly with $x$. Between $x=0$ and $x=0.1$ both $\sigma_{xy}^{{\rm OHE},\uparrow}$ and $\sigma_{xy}^{{\rm OHE},\downarrow}$ and consequently also $R^{\rm top}_{yx}$ (see Eq.~(3) in the main text) change sign. Additionally, the magnitude of $\sigma_{xy}^{{\rm OHE},\uparrow}$ is reduced considerably. Thus, we can qualitatively explain the doping dependence of the THE shown in Fig.~3 of the main text by the doping-dependence of $\sigma_{xy}^{{\rm OHE},s}$.

\begin{figure*}
\includegraphics[width=1.0\linewidth]{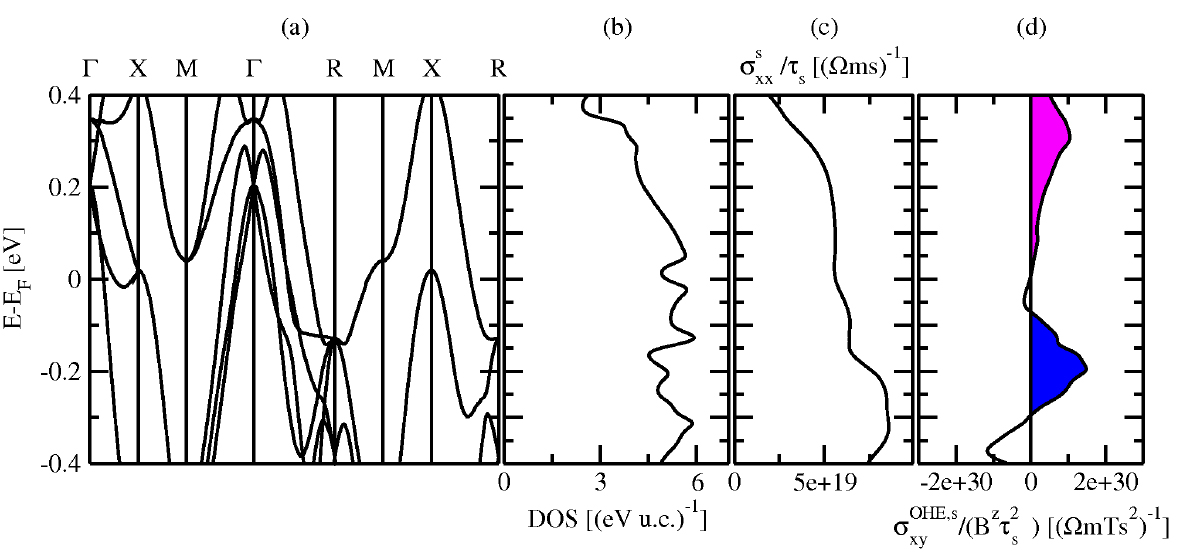}
\caption{Summary of calculated properties of paramagnetic MnSi. (a) Band structure, (b) density of states, (c) normal conductivity and, (d), OHE conductivity. The two salient features of the OHE conductivity are shown in blue and magenta. In magnetic MnSi minority states are shifted up in energy and majority states are shifted down, thereby both the blue and the magenta peaks are shifted towards the Fermi energy. In turn Fig.~\ref{fig_ohe_change_by_doping} may be understood qualitatively in terms of the electronic structure of paramagnetic MnSi. Notably, the large positive $R^{\rm top}_{yx}$ in pure MnSi (in the zero temperature limit) is due to the dominance of the blue peak over the magenta peak. For increasing doping $x$ in Mn$_{1-x}$Fe$_{x}$Si the exchange splitting is reduced leading to the reduction of $R^{\rm top}_{yx}$ shown in Fig.~3~(d) of the main text. 
}
\label{fig_transport_vs_bandstructure}
\end{figure*}

Since the $x$-dependence of the spin-resolved DOS shown in Fig.~\ref{fig_dos_change_by_doping} can be traced to a shift of the Fermi level combined with a decrease of the exchange splitting, one can expect that the qualitative dependence of the THE can be obtained from the electronic structure of paramagnetic MnSi. Thus, we performed an additional calculation of  paramagnetic MnSi (spin-orbit coupling not included). Fig.~\ref{fig_transport_vs_bandstructure} shows the density of states, the normal conductivity and the OHE conductivity as a function of energy, as well as the band structure. Two salient features are identified that are shown in blue and magenta in the figure. In magnetic MnSi minority states are shifted up in energy and majority states are shifted down, thereby both the blue and the magenta peaks are shifted towards the Fermi energy. 
In this way, we can qualitatively understand Fig.~\ref{fig_ohe_change_by_doping} from the electronic structure of paramagnetic MnSi. Thus, according to Eq.~(3), we can trace the large positive $R^{\rm top}_{yx}$ in pure MnSi (in the zero temperature limit) to the dominance of the blue peak over the magenta peak. For increasing doping $x$ in Mn$_{1-x}$Fe$_{x}$Si the exchange splitting is reduced leading to the reduction of $R^{\rm top}_{yx}$ shown in Fig.~3~(d). 

